\documentclass[aps,prd,twocolumn,twoside,superscriptaddress]{revtex4}
\usepackage{physics}
\usepackage{graphicx}
\usepackage{amsmath}
\usepackage{float}
\usepackage{amssymb}
\usepackage{placeins}
\usepackage{array}
\usepackage{subfig}
\usepackage{color}
\usepackage{makecell}
\usepackage{mhchem}
\usepackage{siunitx}
\usepackage{threeparttable}
\usepackage{xr}
\newcommand{\comment}[1]{}

\newcommand{\vecauto}[1]{\ensuremath{\vectorbold*{#1}}}
\newcommand{\vQ}{\vecauto{Q}_{\textrm{ex}}}
\renewcommand{\vr}{\vecauto{r}}

\newcommand{\vq}{\vecauto{q}_{\textrm{L}}}
\newcommand{\vp}{\vecauto{p}}
\newcommand{\vA}{\vecauto{A}}

\newcommand{\Hint}{H^{\textrm{int}}}
\newcommand{\vpolar}{\vecauto{\epsilon}_{\vq\lambda}}

\newcommand{\thetaEx}{\theta_{\textrm{ex}}}
\newcommand{\varphiEx}{\varphi_{\textrm{ex}}}
\newcommand{\thetaL}{\theta_{\textrm{L}}}
\newcommand{\varphiL}{\varphi_{\textrm{L}}}
\newcommand{\QEx}{Q_{\textrm{ex}}}

\newcommand{\vM}{\vecauto{M}}

\begin{document}
  \setcounter{page}{1}
  \date{\today}

\title{Dimensionality and Anisotropicity Dependence of Radiative Recombination in Nanostructured
Phosphorene}
\author{Feng Wu}
\affiliation{Department of Chemistry and Biochemistry, University of California Santa Cruz, Santa Cruz, CA, 95064, USA}
\author{Dario Rocca}\affiliation{Universit\'{e} de Lorraine, LPCT, UMR 7019, 54506 Vand\oe{}uvre-l\`{e}s-Nancy, France}
\affiliation{CNRS, LPCT, UMR 7019, 54506 Vand\oe{}uvre-l\`{e}s-Nancy, France}
\author{Yuan Ping\footnote{yuanping@ucsc.edu}}
\affiliation{Department of Chemistry and Biochemistry, University of California Santa Cruz, Santa Cruz, CA, 95064, USA}


  \begin{abstract}
    The interplay between dimensionality and anisotropicity leads to intriguing optoelectronic properties and exciton dynamics in low dimensional semiconductors. In this study we use nanostructured phosphorene as a prototypical example to unfold such complex physics and develop a general first-principles framework to study exciton dynamics in low dimensional systems.
   Specifically we derived the radiative lifetime and light emission intensity from 2D to 0D systems based on many body perturbation theory, and investigated the dimensionality and anisotropicity effects on radiative recombination lifetime both at 0 K and finite temperature, as well as polarization and angle dependence of emitted light. 
   We show that the radiative lifetime at 0 K increases by an order of $10^3$ with the lowering of one dimension (i.e. from 2D to 1D nanoribbons or from 1D to 0D quantum dots).
   We also show that obtaining the radiative lifetime at finite temperature requires accurate exciton dispersion beyond the effective mass approximation. Finally, we demonstrate that monolayer phosphorene and its nanostructures always emit linearly polarized light consistent with experimental observations, 
   different from in-plane isotropic 2D materials like \ce{MoS2} and h-BN that can emit light with arbitrary polarization, 
   which may have important implications for quantum information applications. 
   \end{abstract}

\maketitle   
Two-dimensional (2D) materials are known for their unprecedented potential in ultrathin electronics, photonics, spintronics and valleytronics applications\cite{Liu2017,Youngblood2015,Shiue2017,Guo2016,Bernardi2017,Eda2013,Britnell2013,Wong2017,Han2016,Feng2017,Avsar2017}. Strong light-matter interactions and atomically-thin thickness lead to exotic physical properties which do not exist in traditional 3D semiconductors. In particular, the monolayer black phosphorus (MBP), commonly known as phosphorene, 
has recently attracted significant interest due to its emerging optoelectronic applications and the development of large scale fabrication methods\cite{Cupo2017,Akhtar2017,Carvalho2016}. 

Unlike other common 2D materials such as graphene, $h$-BN, and transition metal dichalcogenides,
MBP has a strong in-plane anisotropic behavior along two distinctive directions, denoted as ``armchair''and ``zigzag''~\cite{Cupo2017,Pang2017,Li2017,Tran2014,Qiu2017}. For example, the lowest exciton transition is only bright when light is polarized along the armchair direction and the corresponding excitonic wavefunction is much more extended along this direction, leading to so-called ``quasi-1D" excitons\cite{Tran2014}. 
The complex nature of excited states and the anisotropic behavior of excitonic wavefunctions may lead to intriguing effects on the exciton dynamics in MBP, which
are distinctive from the in-plane isotropic 2D systems as we will discuss later.
Further lowering dimensionality is expected to
show an interplay between quantum confinement and quasi-1D excitonic nature which affects its optoelectronic properties in a complex manner. However, because lower dimensional MBP nanostructures (such as nanoribbons and quantum dots) are more difficult to stabilize than two-dimensional MBP\cite{Castellanos-Gomez2014,Yang2015}, the optical measurements and determination of excited state lifetime of these nanostructures are more challenging. Only the exciton recombination lifetime of 2D MBP has been determined to be 211 ps through time-resolved photoluminescence measurements with light
 polarized along the armchair direction at -$10^\circ$C\cite{Yang2015}.
This calls for theoretical studies to predict exciton dynamics of low dimensional MBP nanostructures. 

Previously, the radiative lifetime of excitons of typical in-plane isotropic 2D systems have been computed by coupling Fermi's golden rule with model Hamiltonians or the Bethe-Salpeter equation~\cite{Palummo2015,Chen2018,San-Jose2016,Ayari2018}. However, the radiative lifetime and light emission intensity 
of \textit{in-plane anisotropic systems} (e.g. phosphorene), which have unique dependence on the polarization direction of light, have not been investigated in-depth. Additional outstanding questions involve the dimensionality (i.e. going from the 2D MBP to 1D and 0D nanostructures) and temperature dependence of excited state lifetime. In particular, temperature effects are determined by the exciton dispersion in momentum space,
where the typically used effective mass approximation may break down for low dimensional systems~\cite{Cudazzo2016}.
By answering the above questions, we will propose general principles and pathways of engineering radiative lifetime in anisotropic low-dimensional systems.

The rate of emission of a photon with specific wave-vector $\vq$ and polarization direction $\lambda$ from an exciton with wave-vector $\vQ$ is 

\begin{align}
    \gamma(\vQ,\vq,\lambda) &= \frac{2\pi}{\hbar}\left|\mel{G,1_{\vq\lambda} }{\Hint}{S(\vQ),0}\right|^2\nonumber\\
    &\times \delta(E(\vQ)-\hbar cq_L), \label{eq:decay-general-Qqp}
\end{align}
where $0$ and $1_{\lambda\vq}$ denote absence and presence of a photon, respectively; $G$ is the ground-state; $S(\vQ)$ is the exciton state; $E(\vQ)$ is the exciton energy and $V$ is the system volume. 
  
Two important quantities can be derived from $\gamma(\vQ,\vq,\lambda)$. One is the radiative decay rate $\gamma(\vQ)$ of the exciton that only depends on exciton wavevector $\vQ$; the other is the light emission intensity $I(\vq,\lambda)$, which only depends on photon wavevector $\vq$ and polarization $\lambda$ for a given system. Both quantities can be directly measured from experiments. 

The radiative decay rate of a specific exciton with wave-vector $\vQ$ can be obtained from: 
\begin{align}
    \gamma(\vQ) &= \sum_{\vq\lambda=1,2}\gamma(\vQ,\vq,\lambda); \label{eq:decay-general-Q}
\end{align}
the corresponding lifetime can be simply computed as the inverse of $\gamma(\vQ)$.

By using the dipole approximation and the relation $\vp=-m\frac{i}{\hbar}[\vr,H]$, the exciton transition matrix element in Eq.~\ref{eq:decay-general-Qqp} can be written explicitly as

\begin{align}
    &\mel{G,1_{\vq\lambda}}{\Hint}{S(\vQ),0} \nonumber\\
    =& \sqrt{\frac{e^2 2\pi}{m^2 cV\hbar}\frac{1}{q}}\vpolar\cdot\mel{G}{\vr}{S}. \label{eq:dipole-moment}
\end{align}

The radiative decay rate $\gamma(\vQ)$ can be obtained by combining Eqs.~\ref{eq:decay-general-Qqp}-\ref{eq:dipole-moment}. The exciton energy
$E(\vQ)$ and exciton dipole moments $\mel{G}{\vr}{S}$ necessary as inputs are computed by solving the Bethe-Salpeter equation\cite{Ping2013}, which accurately takes into account the electron-hole interaction (this is mandatory for 2D semiconductors where large exciton binding energy $>$ 0.5 eV has been observed\cite{Huser2013,Qiu2016}). 

In order to study the dimensionality dependence of radiative lifetime in MBP nanostructures, we derived the general radiative decay rate expressions for anisotropic materials from 2D to 0D; detailed derivations can be found in Supporting Information (SI). In general, the radiative decay rate can be written as 
\begin{align}
\gamma(\vQ) &= \gamma_0 Y(\vQ), \label{eq:rate-general-angle-separated}
\end{align}
where $\gamma_0$, which does not depend on the directions of photon or exciton wavevectors, is the exciton decay rate at $\vQ=0$; the dependence on the wavevector direction is instead contained in $Y(\vQ)$, which satisfies $Y(\vQ=0)=1$.

\begin{figure}[htbp]
    \includegraphics[width=0.95\linewidth]{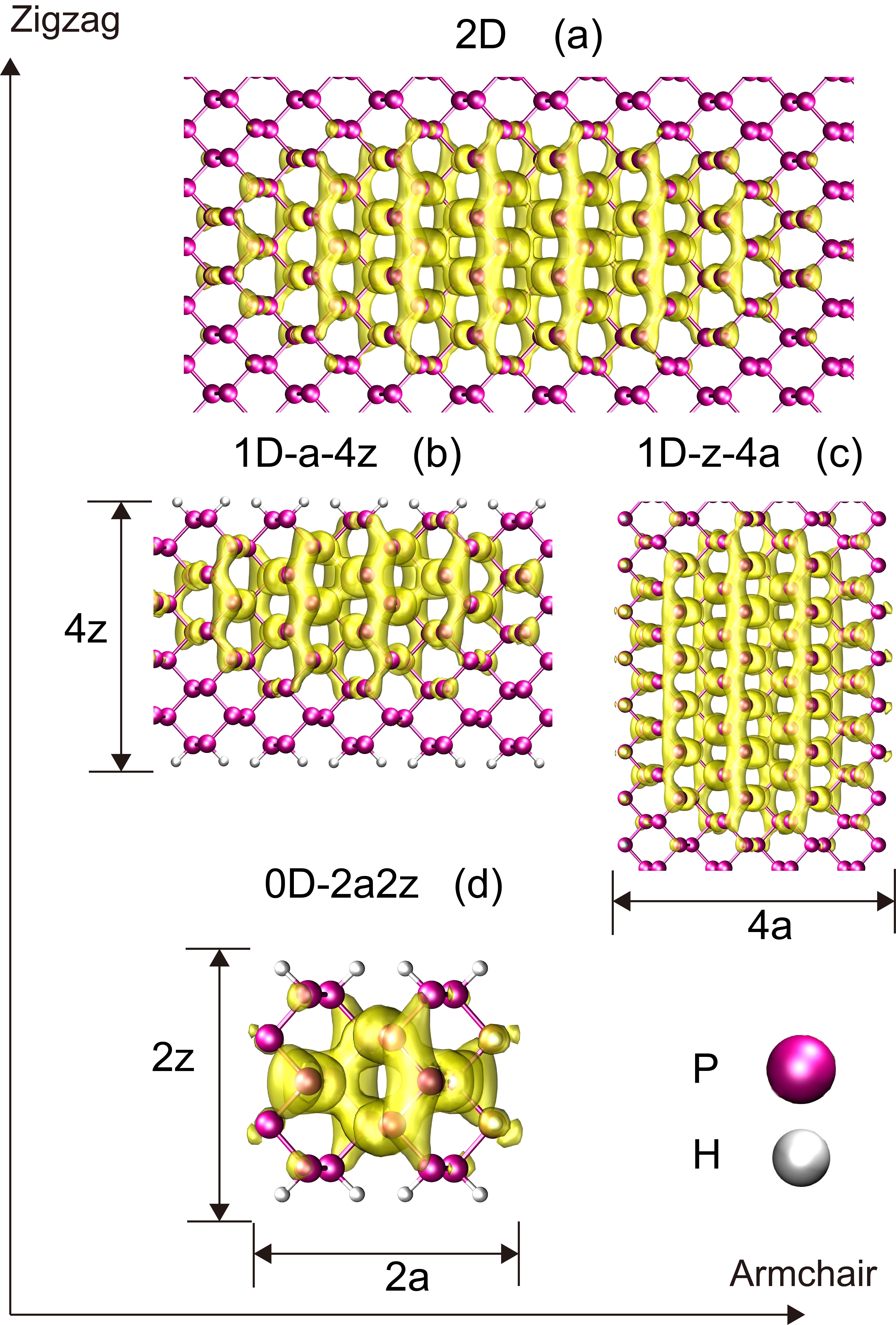}
    \caption{Structural models and exciton wavefunctions of MBP monolayer (a), nanoribbons (b)(c) and quantum dots (d) with dangling bonds terminated by hydrogen atoms. MBP monolayer has two different directions, armchair and zigzag. We constructed two kinds of nanoribbons: ``1D-a-$N$z" where the armchair direction is periodic, and $N$ is the number of unit cells along the non-periodic zigzag direction. ``1D-z-$N$a" where the zigzag direction is periodic, and $N$ is the number of unit cells along the non-periodic armchair direction. Quantum dots are denoted as ``0D-$N$a-$M$z" where $N$ and $M$ give the number of unit cells along the specific directions. The wavefunctions of the lowest exciton of different structures are shown by a yellow isosurface with a value of 0.13 e/\AA$^3$.
    }
    \label{fig:all-structure}
 \end{figure}

To understand the effects of anisotropicity and dimensionality of nanostructured MBP, we computed electronic structure, absorption spectra and radiative lifetimes of various MBP nanostructures from 2D to 0D as shown in Fig.~\ref{fig:all-structure} and Table~\ref{table:exciton-all-lifetime-temperature}. We used open source plane-wave code Quantum-Espresso\cite{QE1,QE2} with Perdue-Burke-Ernzehorf (PBE) exchange-correlation functional~\cite{PBE1}, ONCV norm-conserving pseudopotentials~\cite{ONCV1,ONCV2}. The band structure is computed from GW approximation with the WEST-code~ \cite{WEST,pham2013g}. The absorption spectra and exciton properties are computed by solving the Bethe-Salpeter equation (BSE) in the Yambo-code\cite{YAMBO} for MBP nanoribbons and monolayer systems. For MBP quantum dots, a BSE implementation without explicit empty states and inversion of dielectric matrix is applied instead\cite{PDEPBSE1,PDEPBSE2,PDEPBSE3} to speed up the convergence. More computational details can be found in SI.

\comment{As shown in Fig.~\ref{fig:all-structure} and Table~\ref{table:exciton-all-lifetime-temperature}, we considered MBP nanoribbon (NR) structures\cite{Han2014} with periodicity along the armchair direction (denoted as ``1D-a-$N$z'', where N is the number of unit cells along the non-periodic direction representing the width of the NR) or along the zigzag direction (denoted as ``1D-z-$N$a"). Similarly, the quantum dot structures are denoted as ``0D-$N$a-$M$z", where $N$ and $M$ indicate the number of unit cells along the armchair and zigzag directions, respectively. All the structures considered here have a monolayer thickness. Both MBP nanoribbons and quantum dots' edges have been saturated by hydrogen atoms to remove dangling bonds.}
Optical absorption spectra of the nanostructures considered in this work are provided in SI; here we will summarize the main features that are relevant to the discussion of lifetimes. The first exciton in phosphorene is bright with light polarized along the armchair direction, but dark along the zigzag direction. 
This behavior is inherited by phosphorene nanostructures (nanoribbons and quantum dots). However, for these systems the light absorption and dipole matrix elements along a non-periodic direction are significantly weakened by a ``depolarization effect", which comes from the microscopic electric fields induced by polarization charge in an external field. This effect is included in our absorption spectra calculations by taking into account the local field effects in the Bethe-Salpeter Equation\cite{Marinopoulos2003,Bruno2005}.

{
\setlength{\tabcolsep}{2pt}
\begin{table}[ht]
    \begin{threeparttable}
      \caption{Radiative lifetime of the first exciton of MBP with 2D, 1D and 0D dimensions at 0 K and 263 K, with light polarized along the armchair direction.}
      \label{table:exciton-all-lifetime-temperature}
      \begin{tabular}{cccccc}
        \hline
          System & \thead{Width \\Armchair \\ (\AA)} & \thead{Width \\ Zigzag \\ (\AA)} & \thead{$\hbar\Omega_0$ \\ (eV)}  & \thead{$\mu_A^2$ \\ (a.u.)} & \thead{Lifetime \\ at 0 K }
           \\
        \hline 
        0D-2a2z &    9 &    9 &   2.80 & 0.75 & 57 ns  \\ 
        0D-3a2z &    13 &    9 & 2.39 & 3.49 & 20 ns  \\ 
        0D-4a2z &    18 &    9 & 2.16 & 6.43 & 14 ns  \\ 
        1D-a-2z &    $\infty$ &  9 & 1.73 & 7.42 & 21 ps  \\ 
        1D-a-3z &    $\infty$ &    12 & 1.52 & 9.78 & 21 ps\\ 
        1D-a-4z &    $\infty$  &15 & 1.51  & 11.89 & 17 ps \\ 
        1D-z-2a &    9 &   $\infty$ & 2.72 & 0.04 &  1.0 ns \\ 
        1D-z-3a &    13 &   $\infty$ & 2.40 & 0.45 & 130 ps\\ 
        1D-z-4a &    18 &   $\infty$ & 2.17 & 0.64 & 110 ps \\ 
        2D* &    $\infty$ & $\infty$ & 1.54 & 2.56 & 99 fs  \\ 
        \hline
        2D(Exp)* & & & & & 211 ps\cite{Yang2015} \\
        2D(noEM)* & & & & & 101 ps\\
        2D(EM)* & & & & & 1.8 ns \\
        \hline
      \end{tabular}
      \begin{tablenotes}
          \item * Exciton lifetime measured or computed at 263K. 2D(EM) is computed from effective mass approximation. 2D(noEM) is computed from Eq.~\ref{eq:exciton-disperion-fullform}.
      \end{tablenotes}
  \end{threeparttable}
\end{table}
}

The radiative lifetimes for various MBP nanostructures at $\vQ=0$ (obtained by $1/\gamma_0$) are summarized in Table~\ref{table:exciton-all-lifetime-temperature}. The lifetime of the first exciton at 0 K changes dramatically from 0D ($\approx$ 10-100 ns) to 1D ($\approx$ 10-100 ps) then to 2D ($\approx$ 100 fs), and generally decreases with increasing system sizes. Comparing the radiative decay rate expressions in SI 
we can see that with every decrease of dimension (e.g. from 2D to 1D), an additional $\Omega_0l_i/c$ multiplicative factor appears in the decay rate expression, where $l_i$ is the unit cell size and $\Omega_0$ is the exciton energy at $\vQ=0$.  
This is due to the momentum conservation between photon $q_i$ and exciton $Q_i$ along the periodic direction $i$ and energy conservation ($\hbar c q_L=\hbar\Omega_0$). 

Considering typical values in a semiconductor solid such as $\hbar\Omega_0\approx 0.5-5$ eV and $l=5-10$ Bohr unit cell lattice parameter along one dimension, we can estimate that
\begin{align}
    \Omega_0l / c \approx 10^{-2}-10^{-3}. \label{eq:lifetime-dimension-reduction-term}
\end{align}
Accordingly, each reduction of dimension will give approximately $10^3$ times longer lifetime at 0 K (this qualitative estimation does not take into account the change of dipole matrix elements and exciton energy).  
The above discussion explains the increase of the lifetime by several orders of magnitude when decreasing the dimensionality. 

For systems of the same dimensionality, the lifetime also varies with the size and periodic direction of nanostructures.  
Specifically, the radiative lifetime is inversely proportional to the product of squared dipole matrix element ($\mu_A^2=|\mel{G}{\vr}{S}|^2$) and the exciton energy ($\Omega_0$), i.e. $1/\gamma_0 \propto 1/(\Omega_0^n\mu_A^2)$, where $n$ depends on the dimensionality.
We found that in most cases, by changing the system sizes, the change of $\mu_A^2$ is much larger than that of $\Omega_0$ and dominates the lifetime, as shown in Table~\ref{table:exciton-all-lifetime-temperature}.
For example, the fact that the radiative lifetime of nanoribbons periodic along zigzag (``1D-z") is $10^1-10^2$ times longer than their counterparts periodic along armchair (``1D-a") is directly determined by $\mu_A^2$. Because of the anisotropicity of MBP structure, the only non-zero component of $\mel{G}{\vr}{S}$ of the first exciton is along the armchair direction. In 1D-a nanoribbons the armchair direction is periodic, so the quasi-1D exciton $\mu_A^2$ is similar to 2D. On the other hand, in 1D-z nanoribbons the armchair direction is not periodic 
and $\mu_A^2$ is strongly reduced due to the depolarization effect; however, the dipole moments with light polarized along the armchair direction is still the main contribution to $\mu_A^2$.
This is different from previous work on 1D nanostructures such as carbon nanotubes\cite{Spataru2005} where only the component $\mu_z$ in the periodic direction dominates $\mu_A^2$. Furthermore, 
the quasi-1D nature of the first exciton (extended along the armchair direction and confined along the zigzag in Fig.~\ref{fig:all-structure}) introduces a much larger change in lifetimes as a function of the width for 1D-z nanoribbons compared to 1D-a nanoribbons.

\comment{The quasi-1D nature of the first exciton (extended along the armchair direction and confined along the zigzag in Fig.~\ref{fig:all-structure}) is also important to explain the large changes in the lifetimes as a function of the width for 1D-z nanoribbons. As shown in Table~\ref{table:exciton-all-lifetime-temperature} the square of dipole moments $\mu_A^2$ sharply increases from 0.04 a.u. for the narrowest 1D-z-2a nanoribbon to 0.64 a.u. for the widest 1D-z-4a nanoribbon. As the corresponding exciton energy $\hbar\Omega_0$ presents much smoother changes (10-20$\%$), the lifetime, which is proportional to $1/\mu_A^2$, becomes shorter by increasing the nanoribbon width. On the other hand, the dipole moment and the lifetime change much less when the width of 1D-a nanoribbon increases, due to the confined exciton wavefunction along this direction. }

The radiative lifetimes of 0D systems are significantly longer with respect to the other nanostructures, ranging from $2.0\times10^4$ ps to $1.2\times 10^5$ ps due to the dimensionality dependent factor for decay rates in Eq.~\ref{eq:lifetime-dimension-reduction-term} (see Table~\ref{table:exciton-all-lifetime-temperature}). Similarly to the 1D-z nanoribbons, the 0D-$N$a2z ($N$=2,3,4) series shows a clear trend of lifetime decreasing by increasing the quantum dot size along the armchair direction. 

Next we will discuss the exciton recombination lifetime at finite temperature for 2D phosphorene in order to compare with the experimental results.
The radiative lifetime at finite temperature $T$ is computed by assuming that the recombination process is slow enough to allow excitons at different $\vQ$ to reach thermal equilibrium~\cite{Spataru2005}. The temperature-dependent radiative lifetime is then written as a thermal statistical average:
\cite{Spataru2005,Palummo2015}
\begin{align}
    \gamma(T) &\approx Z^{-1}\int_{\QEx<E_0/\hbar c}\dd \vQ \gamma(\vQ)  \label{eq:rate-temperature-Z} \\
    Z &= \int \dd \vQ e^{-(E(\vQ)-E_0)/k_BT}, \label{partition}
\end{align}
where $E_0=E(\vQ=0)$, and $Z$ is the partition function and the condition $\QEx<E_0/\hbar c$ is imposed by energy conservation. The evaluation of Eqs.~\ref{eq:rate-temperature-Z}-\ref{partition} requires the knowledge of the exciton energy $E$ as a function of $\vQ$. 
 
In previous work~\cite{Spataru2005,Palummo2015,Chen2018} the effective mass approximation for exciton dispersion $E(\vQ)=E(\vQ=0)+\hbar^2\QEx^2/2m$ has been used.
However, the effective mass approximation works best for Wannier-Mott excitons in 3D bulk material. In low dimensional materials such as phosphorene, the weak dielectric screening and delocalized  wavefunctions in plane introduce strong long-range electron-hole exchange interactions. Thus the exciton dispersion violates the parabolic relation for small values of $\QEx$~\cite{Cudazzo2015,Cudazzo2016,Koskelo2017}. 
This calls for a more sophisticated approach to deal with the thermal distribution of excitons beyond the effective mass approximation. 

Following Ref.~\citenum{Cudazzo2016}, we fit $E(\vQ)$ along the armchair and zigzag directions of phosphorene independently using an equation based on the combination of localized and delocalized exciton models:
\begin{align}
    E_i(Q_{\textrm{ex},i}) =& E_0 + \frac{Q_{\textrm{ex},i}^2}{2m_i} \nonumber\\
    &+ \frac{4\pi}{d}\mu_i^2\left(1-\frac{1-e^{-Q_{\textrm{ex},i} d}}{Q_{\textrm{ex},i} d}\right),
    \label{eq:exciton-disperion-fullform} 
\end{align}
where $i$ denotes the direction, $m$ is the effective mass, $\mu_i$ is the component of dipole moment matrix along the specific direction $i$, and $d$ is the thickness of the assumed dielectric medium. 
The partition function is then computed numerically from the fitted $E(\vQ)$ assuming that the dispersion along two directions is not coupled:
\begin{align}
    E(\vQ) =& E_0 + (E_x(Q_{\textrm{ex},x})-E_0) \nonumber \\
    &+ (E_y(Q_{\textrm{ex},y})-E_0).
\end{align}

\comment{GO SI: For nanoribbons no $E(\vQ)$ data available, so we turned back to the effective mass approximation where the effective mass is obtained from GW calculation. Because the density of GW $q$-grid limits the accuracy of parabola fitting of $\partial^2E/\partial k^2$, we take the GW renormalization factor at $\Gamma$ point to scale the effective mass computed from DFT bandstructure as\cite{Miyake2013} by neglecting the $\partial\Re\Sigma(k,\omega)/\partial k$:

\begin{align}
    &\frac{m_{DFT}}{m_{GW}}  \nonumber\\
    =&  \left(1 - \frac{\partial\Re\Sigma(k,\omega)}{\partial \omega}\right)^{-1}\left(1+\frac{\dd k}{\dd \varepsilon}\frac{\partial\Re\Sigma(k,\omega)}{\partial k}\right)  \nonumber\\
    \approx& \left(1 - \frac{\partial\Re\Sigma(k,\omega)}{\partial \omega}\right)^{-1}
\end{align}
}

The temperature-dependent radiative lifetimes of phosphorene (specifically at 263 K and 0 K) are listed in Table~\ref{table:exciton-all-lifetime-temperature}. If the traditional approach based on the effective mass (``EM" value in Table~\ref{table:exciton-all-lifetime-temperature}) approximation is used, a lifetime of 1.8 ns is obtained, which overestimates the experimental result by an order of magnitude. If the more complex model in Eq.~\ref{eq:exciton-disperion-fullform} is used (``noEM" value in Table ~\ref{table:exciton-all-lifetime-temperature}), the computed lifetime of phosphorene at 263 K is 123 ps, which is in good agreement with the experimental result 211 ps\cite{Yang2015}. 
The remaining discrepancy (less than a factor of 2) could be due to the use of a Si/\ce{SiO2} substrate in experimental measurements, which introduces an additional dielectric screening in the material compared to our free-standing system. Our findings show the importance of the accurate exciton dispersion $E(\vQ)$ beyond the effective mass approximation. 

After the discussions of radiative decay rates of excitons that provide time-resolved information, we are going to investigate the polarization-resolved and angle-resolved information of the light emitted from the exciton recombinations.
Angle-resolved and polarization-resolved light emission measurements are performed with a detector that collects photon with momentum $\vq$ and polarization $\lambda$ within a small area around a spherical angle. 
The intensity $I(\vq,\lambda)$ emitted from all possible excitons is
\begin{align}
    I(\vq,\lambda) &= \sum_{\vQ}n(\vQ)\gamma(\vQ,\vq,\lambda), \label{eq:intensity-angle}
\end{align}
where $n(\vQ)$ is the occupation number of the exciton state at momentum $\vQ$. 

By integrating over the norm of $\vq$ and keeping only the non-zero $\gamma(\vQ,\vq,\lambda$) in the summation of $\vQ$ in Eq.~\ref{eq:intensity-angle} (based on the momentum conservation between $\vQ$ and $\vq$ along the systems' periodic directions), the light emission intensity along a specific direction of photon wavevector ($\thetaL,\varphiL$; see Fig.1 in SI) and polarization ($\lambda$) can be expressed as follows: 
\begin{align}
    I(\thetaL,\varphiL,\lambda) &= n_0\Gamma_0\left|\vpolar\cdot\vM \right|^2 \label{eq:rate-q-alldim-oneQ}
\end{align}
where $\vM$ is the normalized exciton dipole moment.

We consider intensity along two polarization directions:
\begin{align}
    I(\thetaL,\varphiL,\textrm{IP}) &= n_0\Gamma_0\left(-M_x\sin \varphiL+ M_y\cos\varphiL\right)^2 \label{eq:intensity-ip} \\
    I(\thetaL,\varphiL,\textrm{OOP})&= n_0\Gamma_0(-M_x\cos\thetaL\cos\varphiL \nonumber\\
       & -M_y\cos\thetaL\sin\varphiL + M_z\sin\thetaL)^2,
       \label{eq:intensity-oop}
\end{align}
where $\textrm{IP}$ (in-plane) and $\textrm{OOP}$ (out-of-plane) denote two polarization directions that are both perpendicular to the photon wavevector $\vq$
(see Fig.1 in SI), and $M_i=\mu_i/\mu_A$ representing the components of the normalized dipole moment along the directions $i=x,y,z$. 
The formulation in Eqs.~\ref{eq:rate-q-alldim-oneQ}-\ref{eq:intensity-oop} is general and only the angle independent term $\Gamma_0$ describes the difference among 2D, 1D and 0D nanostructures (detailed derivations can be found in SI).  

The emitted photon polarization can be described by Stokes parameters\cite{Optics4Hecht,McMaster1954,McMaster1961,Peach2009} $(S_0,S_1,S_2,S_3)$, 
which can be written explicitly as the following: 
\begin{align}
    S_0 &= I(\thetaL,\varphiL,\textrm{IP}) + I(\thetaL,\varphiL,\textrm{OOP}) ,\nonumber\\
    S_1 &= 2I(\thetaL,\varphiL,\textrm{IP}) - S_0 ,\nonumber\\
    S_2 &= 2I(\thetaL,\varphiL,\ang{45}) - S_0 ,\nonumber\\
    S_3 &= 2I(\thetaL,\varphiL,L) -S_0 ,\label{eq:stokes-I}
\end{align}
where $\lambda=\ang{45}$ bisects the angle between IP and OOP, and $\lambda=L$ denotes the left circularly polarized light. 
When $|S_3/S_0|=0$ (1), the light is fully linearly polarized (circularly polarized). By evaluating Eq.~\ref{eq:stokes-I} with inputs from Eqs.~\ref{eq:rate-q-alldim-oneQ}-\ref{eq:intensity-oop}, we can get the Stokes parameters as a function of normalized exciton dipole moments $\vM$. 

We found that because of the in-plane anisotropicity in MBP and its nanostructures, the emitted light is \textit{completely linearly polarized}. As we can consider $M_z\approx0$ (where z is along the perpendicular direction to the material plane), the circular polarization component $S_3$ in  Eq.~\ref{eq:stokes-I} reduces to the following form (detailed derivations can be found in SI):
\begin{align}
    S_3 = n_0\Gamma_0 i\cos\thetaL (M_x^*M_y-M_xM_y^*). \label{eq:stokes-s3}
\end{align}
When one of the following conditions is fulfilled: $M_x=0$, $M_y=0$, or $M_xM_y^*$ is non-zero but real, $S_3$ is zero and the light is fully linearly polarized. Note that the specific choice of the $x,y,z$ directions does not change the conclusion obtained from Eq.~\ref{eq:stokes-s3}, i.e. the polarization of the emitted light\cite{Optics4Hecht}. In the case of anisotropic systems such as MBP nanostructures, by choosing the
$y$ axis along the dark transition direction and the $x$ axis along the bright transition direction (namely $\vM=(1,0,0)$), from Eq.~\ref{eq:stokes-s3} it can be immediately deduced that the emitted light  is linearly polarized. 

In contrast, in-plane isotropic 2D systems with valley degeneracy like \ce{MoS2} may have possible normalized dipole moments $\vM$ that give other types of polarization (circular or elliptical) due to the mixed exciton states between $K$ and $K'$ valleys. For example, when the exciton has $\vM=1/\sqrt{2}(1,i,0)$, 
we obtain $S_3/S_0=1$, and the light is fully circularly polarized.
Mixed exciton states can go through a fast decoherence process that leads to a significant loss of polarization\cite{Chen2018}. However, such decoherence mechanism does not exist in MBP due to its anisotropicity which may be advantageous for quantum information applications that require long coherence time of quantum states. 
The above conclusions agree with the experimental photoluminescence measurements of the first exciton in MBP\cite{Qiao2014, Wang2015, Li2017}, which show a nearly perfect linear polarization. 

In addition, for anisotropic systems like the MBP nanostructures we found that the light emission intensity ($I/(n(\vQ)\Gamma_0$)) 
always has a $\cos^2(\varphi_L)$ angle dependence, where $\varphi_L$ is the azimuth angle of photon wavevector. This is the case regardless of the dimensionality and 
the polarization of the laser that excites the system. This behavior is different from in-plane isotropic 2D systems like BN and MoS$_2$ 
that can have spherical symmetric or $\cos^2(\varphi_L)$ angle dependence based on the mixture between $K$ and $K'$ valleys (polar plots of photoluminescence for different 2D systems can be found in SI). 


In conclusion, we presented a general framework to study exciton radiative lifetimes and light emission intensities in 2D, 1D, and 0D systems from many body perturbation theory. Based on it, important insights were provided on dimensionality and anisotropicity effects on exciton dynamics in phosporene nanostructures.  
For each dimensionality reduction the energy and momentum conservation leads to 
an additional factor of the order of $10^{-3}$ in the decay rates. This explained the general experimental observations that the exciton radiative lifetime is much longer in lower dimensional systems. Furthermore, in order to obtain the accurate radiative lifetime at finite temperature and quantitatively compare with experiments, accurate exciton energy dispersion in momentum space beyond the effective mass approximation must be considered. 
Finally, we demonstrated that the exciton anisotropicity in MBP nanostructures always leads to linearly polarized light emission 
unlike in-plane isotropic 2D materials such as \ce{MoS2} and BN, which can emit light with arbitrary polarization.

\section*{Acknowledgement} 
We thank Francesco Sottile and Matteo Gatti for helpful discussions. This work is supported by NSF award DMR-1760260 and Hellman Fellowship. D.R. acknowledges financial support from Agence Nationale de la Recherche (France) under Grant No. ANR-15-CE29-0003-01. This research used resources of the Center for Functional Nanomaterials, which is a US DOE Office of Science Facility, and the Scientific Data and Computing center, a component of the Computational Science Initiative, at Brookhaven National Laboratory under Contract No. DE-SC0012704, the National Energy Research Scientific Computing Center (NERSC), a DOE Office of Science User Facility supported by the Office of Science of the US Department of Energy under Contract No. DEAC02-05CH11231, the Extreme Science and Engineering Discovery Environment (XSEDE) through allocation TG-DMR160106, which is supported by National Science Foundation Grant No. ACI-1548562\cite{xsede}. 



\end{document}